# Charge Fractionalization in Artificial Tomonaga-Luttinger Liquids with Controlled Interaction Strength

Paul Brasseur[1], Ngoc Han Tu[1], Yoshiaki Sekine[1], Koji Muraki[1], Masayuki Hashisaka[2], Toshimasa Fujisawa[2], and Norio Kumada[1]

[1] *NTT Basic Research Laboratories, NTT Corporation, 3-1 Morinosato-Wakamiya, Atsugi, 243-0198, Japan*
[2] *Department of Physics, Tokyo Institute of Technology, 2-12-1 Ookayama, Meguro, Tokyo 152-8551, Japan*

We investigate charge fractionalizations in artificial Tomonaga-Luttinger liquids (TLLs) composed of two capacitively coupled quantum Hall edge channels (ECs) in graphene. The interaction strength of the artificial TLLs can be controlled through distance $W$ between the ECs. We show that the fractionalization ratio $r$ and the TLL mode velocity $v$ vary with $W$. The experimentally obtained relation between $v$ and $r$ follows a unique function predicted by the TLL theory. We also show that charged wavepackets are reflected back and forth multiple times at both ends of the TLL region.

Many body physics in one dimension (1D) is a challenging field in condensed matter research [1]. In 1D systems, Coulomb interaction strongly modifies electronic states; therefore, the Fermi liquid assumption that interacting fermions in the energy levels close to the Fermi level behave like non-interacting quasiparticles is no longer valid. Instead, 1D systems can be described by Tomonaga-Luttinger liquid (TLL) [2,3], where interacting fermions behave like non-interacting spin and charge density waves with renormalized velocity. The consistency of the TLL model was first recognized through the observation of the power-law dependent suppression of electron tunneling in 1D systems [4-7]. Afterwards, spin-charge separation—the phenomenon where spins and charges are carried by collective waves with different velocities—was probed by momentum-resolved tunneling spectroscopy [8,9]. Three-terminal momentum-resolved tunneling spectroscopy revealed another manifestation of TLL, that is, charge fractionalization [10], where an electron injected from a non-interacting lead in TLL breaks up into left- and right-moving collective waves carrying a fraction of charges. Although these experiments demonstrated characteristic phenomena predicted by the TLL theory, for a complete understanding, a quantitative investigation of interaction effects on the TLL properties is needed.

Recently, the velocity [11-13] and charge fractions [14,15] have been directly measured using artificial TLLs composed of two interacting edge channels (ECs) in a quantum Hall (QH) state, where an EC is a chiral 1D channel propagating along the periphery of a two-dimensional QH system. These measurements became possible by addressing ECs with different spins or propagation directions separately [16,17]. In such systems, tuning of the interaction strength is also possible by changing the distance between the ECs [15,18]. Combining these features enables us to investigate the relation among interaction strength, velocity, and charge fractions in TLLs.

Here, we investigate the charge fractionalization process in artificial TLLs with controlled interaction strength. To change the interaction strength over a wide range, we used ECs separated by a narrow etched line in graphene. Using a narrow etching instead of a narrow gate and graphene instead of a GaAs/AlGaAs heterostructure is important for achieving a strong inter-EC interaction. This is because a gate partially screens the coupling [14,15] and the minimum inter-EC distance in a GaAs/AlGaAs heterostructure is limited by the presence of the depletion layer on both sides of the etched line. Furthermore, the sharp edge potential and resultant narrow ECs in graphene are useful for avoiding edge reconstructions [19]. We prepared several samples with different etching line width $W$. By time-resolved transport measurements, we show that a wavepacket with charge $Q$ injected in the TLL region is fractionalized into wavepackets with charge $rQ$ and $(1-r)Q$. We obtained the fractionalization ratio $r$ and velocity in the TLL region for several values of $W$. The experimentally obtained relation between $r$ and velocity follows a simple analytical function with the information about the dielectric environment and geometry included only in the proportionality factor. These results verify the TLL model over a wide parameter range. We also show that charged wavepackets are reflected back and forth multiple times at both ends of the TLL region.

The graphene was grown by thermally decomposing a $6H$-SiC(0001) substrate. The carriers are electrons, and the density is about $5 \times 10^{11}$ cm$^{-2}$. The mobility is about 12000 cm$^2$/Vs. After Cr/Au ohmic contacts had been deposited, the surface of the graphene was covered with 100-nm-thick hydrogen silsesquioxane (HQS) and 60-nm-thick SiO$_2$ insulating layers. As shown in Fig. 1(a), the graphene is separated into two parts by etching with width $W$ and length $L$. We measured several samples with different combinations of $W = 0.3, 2, 10,$ and $50$ μm and $L = 180, 350,$ and $605$ μm. A perpendicular magnetic field is applied from the front of the sample so that the chirality of ECs becomes counterclockwise. TLL composed of a pair of counter-propagation ECs is formed around the etched region. The time-domain measurement of the charge fractionalization process is based on the excitation of a charged wavepacket and the detection of the time-dependent current. The charged wavepacket with a rough width of $\sim 200$ μm (see supplemental material for details) is excited by a voltage pulse applied to the injection gate that has $10 \times 10$ μm$^2$ overlap with graphene. It propagates as edge magnetoplasmons (EMPs) in ECs [20-23] and undergoes charge fractionalizations in the TLL region. The fractionalized wavepackets are detected through two ohmic contacts, Det1 and Det2, which are connected to each EC at the downstream of the TLL region. In order to reject direct crosstalk between the injector and detectors, we took the difference between the detector currents obtained at two DC bias voltages on the injector. The bias modulates the amount of excited charge, and thus the amplitude of the main signal, while retaining the crosstalk amplitude [24] (see supplemental material for details). Measurements were performed in a 10-T magnetic field at 1.5 K, where pronounced EMP transport in the QH state at Landau level filling factor ν = 2 has been observed in the same material [25,26].

Figures 1(b) and (c) show the charge fractionalization processes in the time domain in samples with $W = 2$ μm and the three different $L$ values. An EMP wavepacket with charge $Q$ is injected at time $t = 0$. The first charge fractionalization occurs when the EMP wavepacket reaches the TLL region. In this region, the wavepacket with charge $Q$ propagating in the upper EC induces charge $-rQ$ on the lower EC, and this forms right-moving TLL plasmons carrying charge $(1-r)Q$ [Fig. 1(a)]. Simultaneously, because of the charge conservation on the lower EC, charge $rQ$ is reflected, which is detected as a current peak by Det2 at $t = t_1$ [dashed vertical line in Figs. 1(b) and (c)]. The second charge fractionalization occurs when the TLL plasmons arrive at the right end of the TLL region, where charge $-rQ$ is reflected back to the left with induced charge $r^2Q$, while charge $(1-r^2)Q$ is transmitted and then detected by Det1 at $t = t_2$ [open triangles in Fig. 1(b)]. By means of the same process at the left end, charge $-r(1-r^2)Q$ appears on Det2 at $t = t_3$ [solid triangles in Fig. 1(c)]. Since there is no charge tunneling between the ECs, after subsequent multiple charge fractionalizations, the total charge arriving at Det1 and Det2 should eventually become $Q$ and zero, respectively (supplemental material).

As mentioned above, the amplitudes of the peaks detected at $t = t_2$ by Det1 and at $t = t_1$ by Det2 are $(1-r^2)Q$ and $rQ$, respectively. From the comparison of the amplitudes, $r$ is estimated to be $0.48 \pm 0.04$ for $W = 2$ μm, and the corresponding Luttinger parameter, which represents interaction strength, is $g = \frac{1-r}{1+r} \sim 0.35$ [17]. Time delays $t_2 - t_1$ and $t_3 - t_1$ correspond to the time of flight of TLL plasmons in the TLL region for one way and a round trip, respectively. Indeed, they increase linearly with $L$ and the slope of $t_3 - t_1$ vs $L$ is twice as large as that of $t_2 - t_1$ vs $L$ [inset of Fig. 1(c)]. Note that $t_3 - t_1$ for $L = 180$ μm is overestimated because of the overlap of the current peak and dip at Det2. From the slope, the velocity of TLL plasmons $v = 1.2 \times 10^6$ m/s is obtained. This value is smaller than the reported EMP velocity of $\sim 1.7 \times 10^6$ m/s at $\nu = 2$ in graphene devices with a similar dielectric environment [26]. This indicates that TLL plasmons are slowed down by interactions between the ECs compared to EMPs in an isolated EC.

To investigate the properties of TLL modes over a wide range of interaction strength, we carried out similar measurements on samples with different $W$. Figure 2 shows the results for samples with $W = 0.3, 2, 10,$ and $50$ μm and $L = 605$ μm. The amplitude of the peak and dip at Det2 becomes larger with decreasing $W$. The time of flight, indicated by open and solid triangles, also changes with $W$—it becomes larger as $W$ is decreased. These observations are consistent with the expectation that $r$ increases and $v$ decreases with strengthening of the inter-EC interaction.

From the data in Fig. 2, the values of $r$ and $v$ are extracted for each $W$. Figure 3(a) shows $r$ as a function of $W$. For $W = 0.3$ μm, $r \sim 0.55$ ($g \sim 0.29$) is obtained. This value is more than one order of magnitude larger than that obtained previously [14], demonstrating that the combination of graphene and narrow etching is effective for achieving large $r$. As $W$ is increased, $r$ decreases slowly and becomes $r \sim 0.2$ ($g \sim 0.67$) at $W = 50$ μm. Theoretically, $r$ is determined by the intra- and inter-EC interactions represented by $U$ and $V$, respectively: [17,27]

$$r = (U - \sqrt{U^2 - V^2})/V. \tag{1}$$

It is reasonable to assume that $U = 1.7 \times 10^6$ m/s corresponds to the EMP velocity in an isolated EC and is independent of $W$. On the other hand, $V$ depends on $W$. Assuming that the width of the ECs is much smaller than $W$ and the plasmon wavelength $\lambda$ is larger than $W$, $V \sim \frac{\sigma_{xy}}{2\pi\epsilon^*\epsilon_0} \ln\frac{\lambda}{W}$, where $\sigma_{xy}$ is the Hall conductance and $\epsilon^*$ is the effective dielectric constant. The dashed line in Fig. 3(a) represents Eq. (1), demonstrating that the observed slow decrease in $r$ with $W$ stems from the logarithmic dependence of $V$ on $W$. Here, we used $\epsilon^* = 6.2$ and $\lambda = 200$ μm, which correspond to the average dielectric constant of the environment and the estimated width of the charged wavepacket, respectively.

In Fig. 3(b), $v$ is plotted as a function of $r$. All the measured $v$ values are smaller than $1.7 \times 10^6$ m/s for an isolated EC [26]. As $r$ decreases, $v$ approaches the value for the isolated EC limit at $r = 0$. The behavior of $v$ can be understood by representing $v$ as a function of $U$ and $r$:

$$v = \sqrt{U^2 - V^2} = U(1 - r^2)/(1 + r^2). \tag{2}$$

This indicates that the details of the dielectric environment and geometry are included only in the proportionality factor $U$, and that once $U$ is experimentally obtained, the TLL mode velocity is uniquely determined by its effective charge, irrespective of the dielectric environment and geometry of the 1D channels. Experimentally obtained $v$ vs $r$ is well reproduced by the trace based on Eq. (2) without any adjustable parameters [dashed line in Fig. 3(b)]. These results quantitatively verify the TLL model over a wide range of interaction strength. Furthermore, the results indicate that the system with a large length scale of $W = 50$ μm can be described by TLL. Since 50 μm is as large as the typical width of a Hall bar device, this implies that interaction between ECs on opposite sides of a Hall bar is not negligible [28].

Finally, we discuss the current waveform in more detail while taking multiple charge fractionalization processes into account. Multiple charge fractionalization processes yield a series of current pulses with charge $(1 - r^2)Q, r^2(1 - r^2)Q, r^4(1 - r^2)Q, \ldots$ on Det1 and $rQ, -r(1 - r^2)Q, -r^3(1 - r^2)Q, \ldots$ on Det2 at a constant time interval [inset of Fig. 4(b)]. When $r$ is as large as 0.5, the terms with higher order than $r^2$ are not negligible and multiple fractionalization processes should be detectable. In Fig. 4, to examine the effect of higher order processes, we replot the data for the sample with $W = 0.3$ μm and $L = 605$ μm together with the calculated current for $r = 0.5$ [29]. In the calculation, we assumed that the propagation properties of TLL plasmons are the same as those of EMPs except for the coefficient $(1 - r^2)/(1 + r^2)$ of the velocity [Eq. (2)], and used the parameters for the dispersion and dissipation determined for EMPs [26]. Then the current waveform is obtained by adding the waveform for each fractionalization process with the coefficient corresponding to the expected charge amount. Comparison between the measured and calculated currents suggests that the small bump after the main peak and the slowly decaying tail in the measured current stem from multiple charge fractionalization processes.

In summary, we studied charge fractionalization processes in artificial TLL systems composed of two closely separated counter-propagating ECs. We prepared such systems in graphene using a narrow line etching. A large fractionalization ratio $r \sim 0.55$ was achieved for the narrowest etching $W = 0.3$ μm. We demonstrated that $r$ can be changed by a factor of two by changing $W$ between 0.3 and 50 μm. The velocity of TLL plasmons measured at the same time also changes with $W$. The experimentally obtained relation between $v$ and $r$ follows an analytical function without any adjustable parameters. These results verify the TLL theory over a wide parameter range. The tuning of $r$ and $v$ demonstrated in this work is useful for applications of plasmonic devices [27,30,31].

Acknowledgements
We thank H. Irie and K. Sasaki for valuable discussions and A. Tsukada for sample fabrication. This work was supported by JSPS KAKENHI Grant Number JP15H05854.

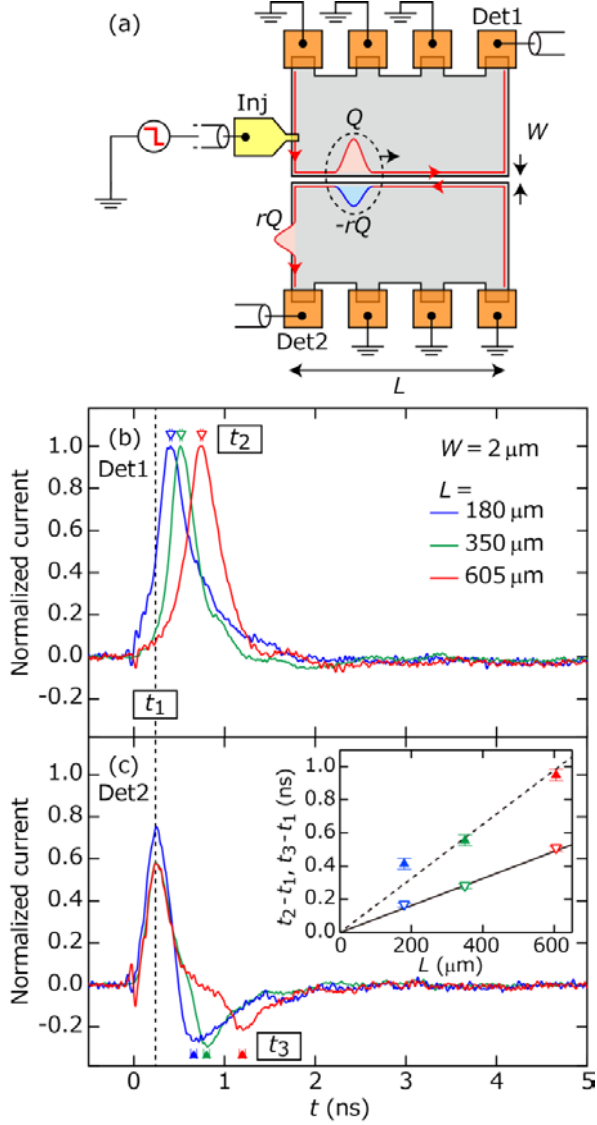

FIG. 1. (color online) Time-resolved measurement of charge fractionalization. (a) Schematic of a sample. An injection gate (yellow) and ohmic contacts (orange) are patterned on graphene (grey). High-frequency lines are connected to the injection gate and two ohmic contacts, labeled Det1 and Det2. (b) and (c) Current as a function of time $t$ at Det1 and Det2, respectively, in three samples with different values of $L$ (180, 350, and 605 μm) at $W = 2$ μm. The plotted current is normalized by the peak amplitude at Det1 to compensate for the sample dependence of $Q$, which stems from fluctuations in the size of the injector and thickness of the gate insulator. The inset shows $t_2 - t_1$ (open triangles) and $t_3 - t_1$ (solid triangles) as a function of $L$. Error bars originate from current fluctuations due to residual crosstalk (Fig. S1 in supplemental material). The solid line is the result of a line fitting. The slope of the dashed line is twice as large as that of the solid line.

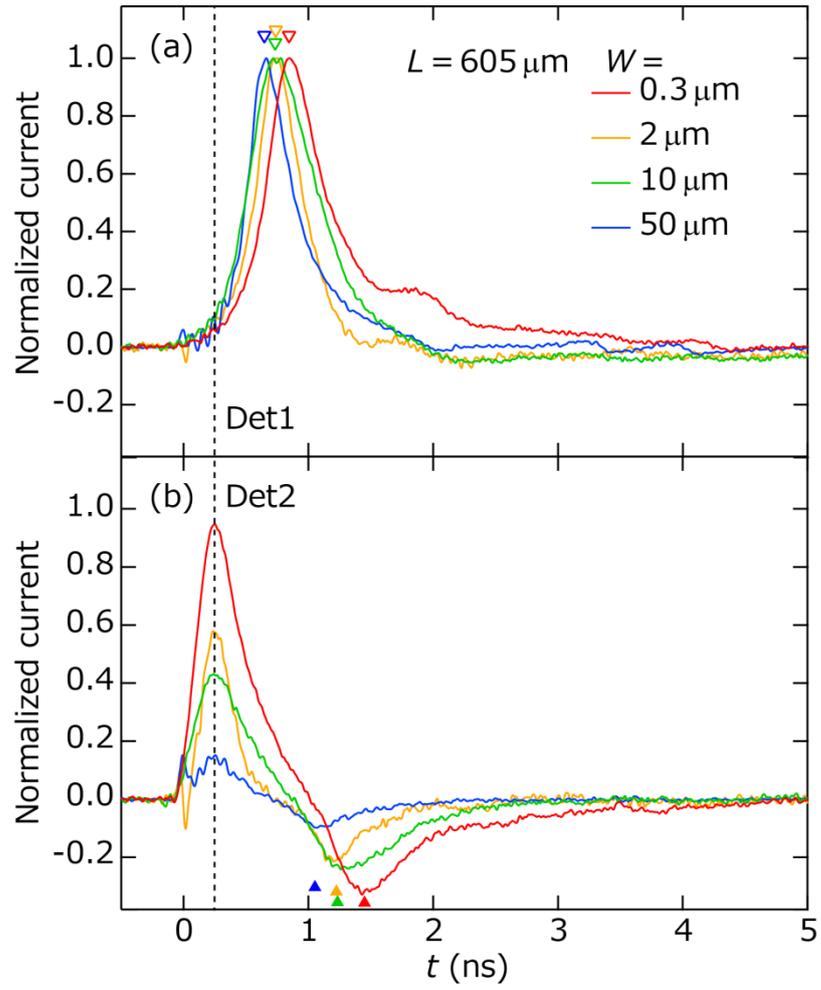

FIG. 2. (color online) Results for different inter-EC couplings. (a) and (b) Current as a function of $t$ at Det1 and Det2, respectively, for four samples with different values of $W$ (0.3, 2, 10, and 50 μm) at $L = 605$ μm. The plotted current is normalized by the peak amplitude at Det1.

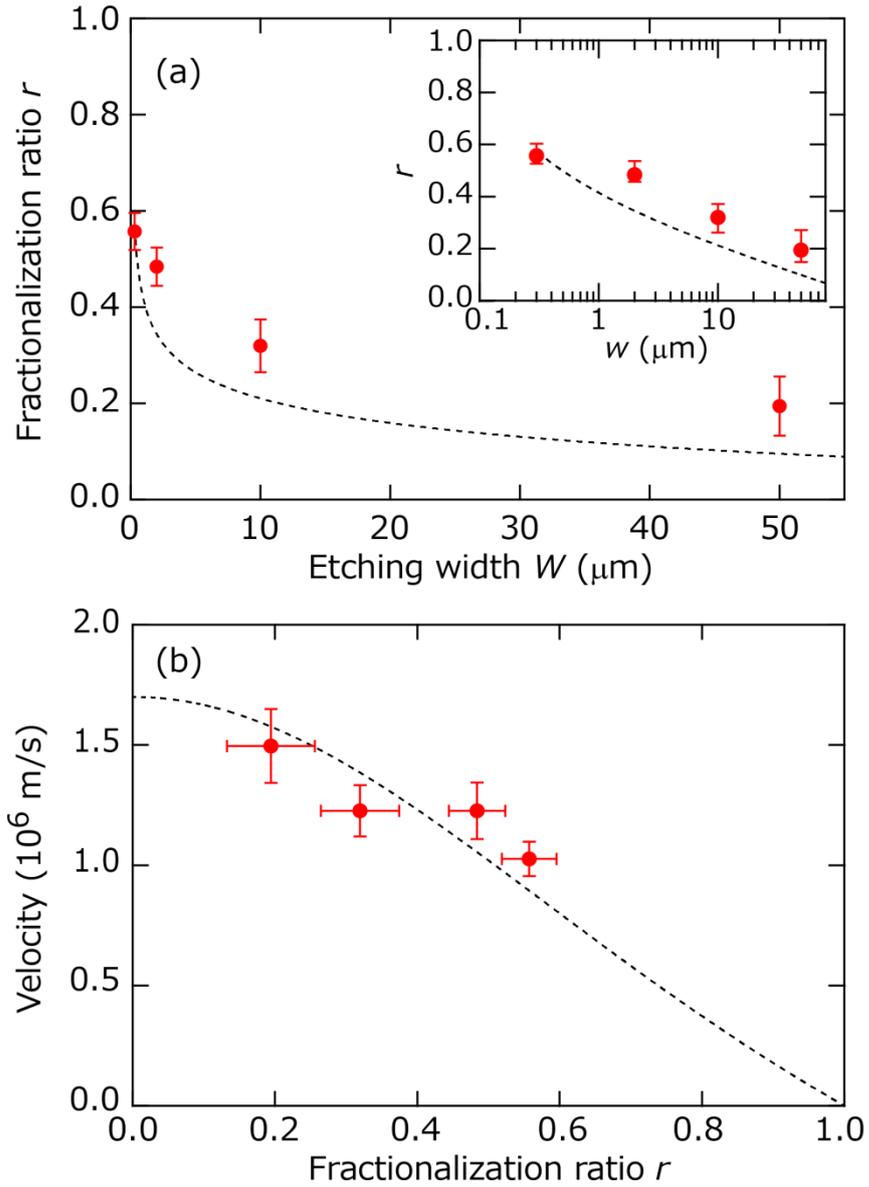

FIG. 3. (color online) (a) $r$ as a function of $W$. The error bars represent fluctuations of $r$ for samples with different $L$. Inset shows $r$ plotted as a function of $W$ in logarithmic scale. (b) TLL plasmon velocity as a function of $r$, extracted from samples with $L = 605$ μm. The size of the error bars for the velocity is determined by the error of the peak position, while that for $r$ is identical to (a). The dashed lines in (a) and (b) represent the theoretical predictions based on Eqs. (1) and (3), respectively.

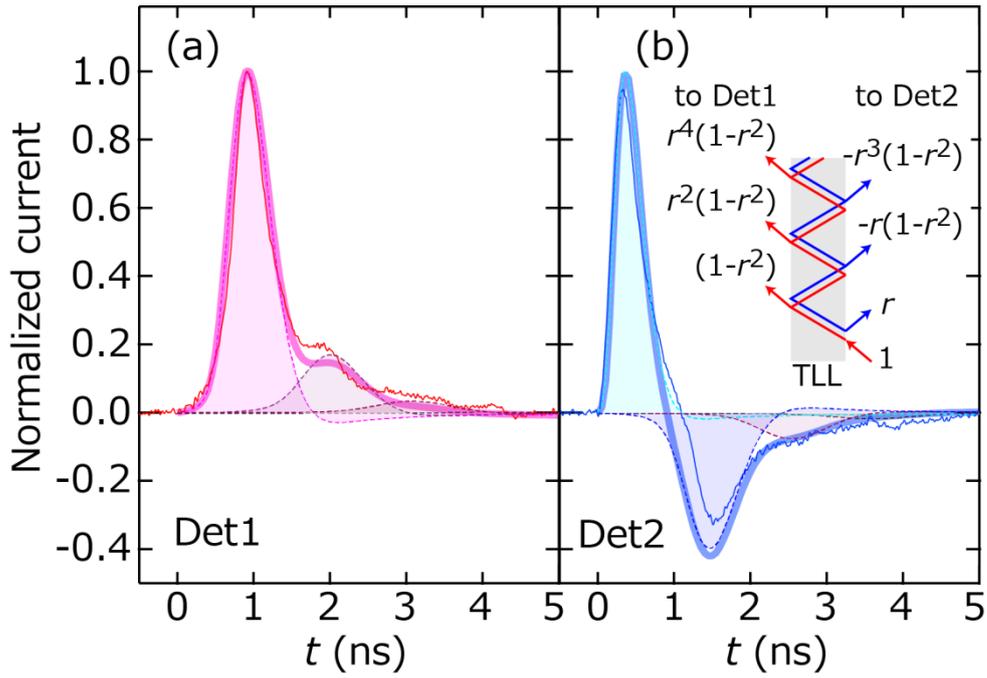

FIG. 4. (color online) Calculated current (thick traces) at (a) Det1 and (b) Det2 obtained by adding current pulses for each fractionalization process (filled areas). Thin solid traces represent the measured current for $W = 0.3$ μm and $L = 605$ μm, which are identical to those in Fig. 2. The inset illustrates the ratio of charge transmitted for each fractionalization process.